# Anomalous Brownian motion via linear Fokker-Planck equations

A. O. Bolivar

## Abstract

According to a traditional point of view Boltzmann entropy is intimately related to linear Fokker-Planck equations (Smoluchowski, Klein-Kramers, and Rayleigh equations) that describe a well-known nonequilibrium phenomenon: (normal) Brownian motion of a particle immersed in a thermal bath. Nevertheless, current researches have claimed that non-Boltzmann entropies (Tsallis and Renyi entropies, for instance) may give rise to anomalous Brownian motion through nonlinear Fokker-Planck equations. The novelty of the present article is to show that anomalous diffusion could be investigated within the framework of non-Markovian linear Fokker-Planck equations. So on the ground of this non-Markovian approach to Brownian motion, we find out anomalous diffusion characterized by the mean square displacement of a free particle and a harmonic oscillator in absence of inertial force as well as the mean square momentum of a free particle in presence of inertial force.

# 1. Introduction

The jittering movement of a Brownian particle of mass $m$, position $x$ and linear momentum $p$ immersed in a given environment is usually described by the Fokker-Planck equation in phase space [1,2]

$$\frac{\partial \mathcal{F}(x,p,t)}{\partial t} = -\frac{p}{m}\frac{\partial \mathcal{F}(x,p,t)}{\partial x} + \frac{\partial}{\partial p}[2\gamma p - f(x)]\mathcal{F}(x,p,t) + D_p \frac{\partial^2 \mathcal{F}(x,p,t)}{\partial p^2}, \quad (1)$$

where the function $\mathcal{F}(x,p,t)$ provides the probability $\mathcal{F}(x,p,t)dxdp$ for that Brownian particle being in the infinitesimal region $dxdp$ around the point $(x,p)$ at time $t$. Hence, $\mathcal{F}(x,p,t)$ is termed the probability distribution function of the various possible values of the positions and momenta of the Brownian particle at time $t$. In Eq. (1) $f(x) = -dV(x)/dx$ is a force derived from the potential $V(x)$ and $2\gamma$ denotes the constant of friction between the Brownian particle and the environmental particles.

If it is assumed the environment to be a thermal reservoir (heat bath) thermodynamically characterized by the Boltzmann constant $k_B$ and the temperature $T$, then, according to the tenets of equilibrium statistical mechanics based on the Boltzmann entropy, the possible values of positions and momenta of the Brownian particle are given by the Maxwell-Boltzmann probability distribution function

$$\mathcal{F}(x,p) = Ne^{-\frac{1}{k_BT}\left[\frac{p^2}{2m}+V(x)\right]}, \quad (2)$$

where $N = 1/\sqrt{2\pi m k_B T}\int_{-\infty}^{\infty} e^{-V(x)/k_BT}dx$ is a normalization constant. Inserting the steady solution (2) into Eq. (1), the diffusion coefficient $D_p$ of the Brownian particle is determined as

$$D_p = 2\gamma m k_B T. \quad (3)$$

The Fokker-Planck equation (1) with the time-independent diffusion coefficient (3) is known as the Klein-Kramers equation [3,4].

The integration of the Klein-Kramers equation (1), with Eq.(3), over the position $x$ leads in the case of a free particle to the Rayleigh equation

$$\frac{\partial \mathcal{F}(p,t)}{\partial t} = -\frac{p}{m}\frac{\partial \mathcal{F}(p,t)}{\partial x} + 2\gamma\frac{\partial}{\partial p}[p\mathcal{F}(p,t)] + 2\gamma m k_B T\frac{\partial^2 \mathcal{F}(p,t)}{\partial p^2}, \quad (4)$$

whose steady solution is the Maxwell distribution

$$\mathcal{F}(p) = \frac{1}{\sqrt{2\pi m k_B T}}e^{-\frac{p^2}{2mk_BT}}. \quad (5)$$

As far as the strong friction condition is concerned, the Brownian motion turns out to be described by the Fokker-Planck equation in configuration space [1,2]

$$\frac{\partial \mathcal{F}(x,t)}{\partial t} = \frac{1}{\gamma m}\frac{\partial}{\partial x}\left[\frac{dV(x)}{dx}\mathcal{F}(x,t)\right] + D_x \frac{\partial^2 \mathcal{F}(x,t)}{\partial x^2}, \qquad (6)$$

where $\mathcal{F}(x,t)$ is the probability distribution function of the values of the positions of the Brownian particle at time $t$. Again, assuming a thermal environment to be characterized by the Boltzmann distribution

$$\mathcal{F}(x) = N' e^{-\frac{V(x)}{k_B T}}, \qquad (7)$$

with $N' = 1/\int_{-\infty}^{\infty} e^{-V(x)/k_B T} dx$, the diffusion coefficient $D_x$ is found to be

$$D_x = \frac{k_B T}{\gamma m}. \qquad (8)$$

The Fokker-Planck equation (6) with the diffusion coefficient (8) is called the Smoluchowski equation [5]. Solving this diffusion equation for a free particle $V(x) = 0$, Einstein [6] found the mean square displacement with the time

$$\langle X^2(t) \rangle = \frac{k_B T}{\gamma m} t, \qquad (9)$$

which characterizes the diffusion process as normal. Langevin [7], in turn, showed that the Einsteinian diffusion law (9) is valid for $t \to \infty$, i.e., $t \gg 1/\gamma m$.

Upshots (1-9) show that there seems to exist a close relationship between Boltzmann entropy, Maxwell-Boltzmann distributions, linear Fokker-Planck equations and normal diffusion [8-15]:

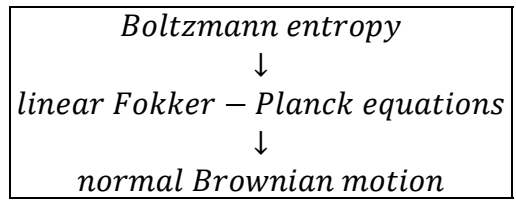

However, in the relatively recent literature on Brownian motion it has been reported that generalized diffusion movements can occur. For instance, as far as disordered environments are concerned, the Einsteinian diffusion law (9) turns out to be no longer valid for certain physical systems, such as dopped conductors and semiconductors, disordered lattice, and porous media [16-19]. In these cases, the long-time behavior of the mean square of a given generic physical quantity $G(t)$ turns out to be governed by the expression [18]

$$\langle G^2(t)\rangle \sim C^{(\lambda)} t^\lambda, \qquad t \to \infty, \tag{10}$$

where $C^{(\lambda)}$ is a constant labeled by the index $\lambda$. Brownian movement characterized by Eq. (10) is called anomalous (or non-Einsteinian) diffusion.

The disordered environment containing impurities, defects, or some sort of intrinsic disorder, gives rise to two different domains of anomalous diffusion according to the values of $\lambda \neq 1$ in Eq. (10) [19]: (a) subdiffusion for $0 < \lambda < 1$; and (b) superdiffusion for $\lambda > 1$. Specific cases of superdiffusion happen for the cases $\lambda = 2$ and $\lambda = 3$ which are said to be ballistic and turbulent, respectively.

The non-Einsteinian diffusion (10) may be derived through generalizations of the linear Fokker-Planck equations (1), (4) and (6), such as fractional Fokker-Planck equations [19-21] or via nonlinear Fokker-Planck equations built up on the basis of non-Boltzmannian entropies (e.g., Tsallis and Renyi entropies [22,23]) [24-30]. In this context of generalized statistical mechanics, there is a strenuous effort to show a close relationship between non-Boltzmann entropies, nonlinear Fokker-Planck equations and anomalous diffusion:

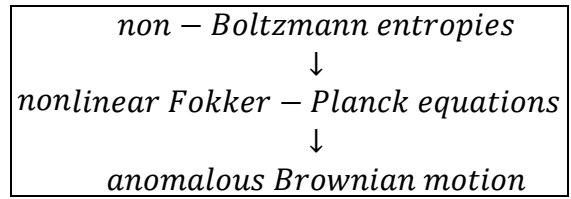

However, our aim in this article is to study the non-Einsteinian diffusion (10) within the framework of non-Markovian linear Fokker-Planck equations already developed elsewhere [31-33]. The paper is then laid out as follows. Both the Klein-Kramers equation (1) and the Smoluchowski equation (6) are generalized in Sects 2 and 3, respectively. The anomalous diffusion phenomenon is addressed in Sec. 4 as much in configuration space as in phase space. Some considerations on the implications of the present approach on the statistical thermodynamics of anomalous diffusion phenomena are pointed out in Sec. 5.

## 2. Generalized Klein-Kramers equation

The movement of a Brownian particle of mass $m$, position $X \equiv X(t)$, and linear momentum $P \equiv P(t) = m dX(t)/dt$ in presence of a generic environment may be described by the following system of stochastic differential equations

$$\frac{dP(t)}{dt} = -\frac{dV(X)}{dX} - \beta P(t) + b\Psi(t), \tag{11a}$$

$$\frac{dX(t)}{dt} = \frac{P(t)}{m}. \tag{11b}$$

The term $dP(t)/dt$, called inertial force, is proportional to the acceleration $d^2X(t)/dt^2$, while $-dV(X)/dX$ is a conservative force derived from the Brownian particle's potential energy $V(X)$. The environmental force $F_{\text{env}}(P,\Psi) = -\beta P + b\Psi(t)$ is made up by the dissipative force $-\beta P$ and the fluctuating force $L(t) = b\Psi(t)$, dubbed Langevin's force. In Eq. (11a) if $\Psi(t)$ has dimensions of time$^{-1/2}$, then the parameter $b$ displays dimensions of mass × length × time$^{-3/2}$.

From a mathematical point of view, both Eqs. (11) are defined within the Kolmogorov probabilistic framework in which $X(t), P(t)$, and $\Psi(t)$ are interpreted as stochastic processes (or random variables) in the sense that there is a probability distribution function, $\mathcal{F}_{XP\Psi}(x,p,\psi,t)$, expressed in terms of the possible values (realizations) $x = \{x_i(t)\}, p = \{p_i(t)\}$, and $\psi = \{\psi_i(t)\}$, with $i \geq 1$, distributed about the sharp values $x', p'$, and $\psi'$ of $X(t), P(t)$, and $\Psi(t)$, respectively. By contrast, the parameters $t$ (time), $\beta$ (frictional constant), $m$ (mass), and $b$ (fluctuation strength) in the stochastic differential equations (11) show up as non-random quantities so having sharp values.

*Kolmogorov equation in phase space.* The stochastic differential equations (11) give rise to the following Kolmogorov equation in phase space (see Appendix A)

$$\frac{\partial \mathcal{F}(x,p,t)}{\partial t} = \mathbb{K}\mathcal{F}(x,p,t) \tag{12}$$

for the (marginal) probability distribution function (the Kolmogorov function)

$$\mathcal{F}(x,p,t) = \int_{-\infty}^{\infty} \mathcal{F}_{XP\Psi}(x,p,\psi,t)\, d\psi. \tag{13}$$

Equation of motion (12) is an exact equation in which the Kolmogorovian operator $\mathbb{K}$ acts on the function $\mathcal{F}(x,p,t)$ according to the prescription (see Appendix A)

$$\mathbb{K}\mathcal{F}(x,p,t) = \sum_{k=1}^{\infty}\sum_{r=0}^{k} \frac{(-1)^k}{r!(k-r)!} \frac{\partial^k}{\partial x^{k-r} \partial p^r}\left[A^{(k-r,r)}(x,p,t)\mathcal{F}(x,p,t)\right], \tag{14}$$

the time-dependent coefficients $A^{(k-r,r)}(x,p,t)$ being given by

$$A^{(k-r,r)}(x,p,t) \equiv \lim_{\epsilon \to 0} \left[ \frac{\langle (\Delta X)^{k-r} \rangle \langle (\Delta P)^r \rangle}{\epsilon} \right], \tag{15}$$

where both average quantities $\langle (\Delta X)^{k-r} \rangle$ and $\langle (\Delta P)^r \rangle$ are deemed to be calculated about their sharp values $x'$ and $p'$, respectively, i.e.,

$$\mathcal{F}_{XP\Psi}(x,p,\psi,t) = \delta(x-x')\delta(p-p')\mathcal{F}_{\Psi}(\psi,t). \tag{16}$$

Both increments $\Delta X \equiv X(t+\epsilon) - X(t)$ and $\Delta P \equiv P(t+\epsilon) - P(t)$ in Eq. (15) are evaluated from Eqs. (11) and hence expressed respectively in the integral form

$$\Delta X = \frac{1}{m} \int_t^{t+\epsilon} P(t)dt, \tag{17}$$

$$\Delta P = -\int_t^{t+\epsilon} \left[ \frac{dV(X)}{dX} + \beta P(t) \right] dt + b \int_t^{t+\epsilon} \Psi(t)dt. \tag{18}$$

Since the Kolmogorov equation (12) depends on an infinite number of autocorrelation functions of the random force $\langle L(t_1)L(t_2)\dots L(t_n) \rangle$, with $n = 2, \dots, \infty$, both Eqs. (11) and (12) describe the time evolution in phase space of a Brownian particle immersed in a generic non-Gaussian environment.

In addition, the non-Gaussian equation of motion (12) is to be solved starting from a given initial condition $\mathcal{F}(x,p,t=0) = \mathcal{F}_0(x,p)$. If the solution $\mathcal{F}(x,p,t)$ renders steady in the long-time regime

$$\lim_{t \to \infty} \mathcal{F}(x,p,t) = \mathcal{F}(x,p), \tag{19}$$

then it is said that the Brownian particle has reached the same stationary state of the environment. Otherwise, i.e., if there exists the asymptotic

$$\lim_{t \to \infty} \mathcal{F}(x,p,t) \approx \tilde{\mathcal{F}}(x,p,t), \tag{20}$$

the Brownian particle holds in a non-stationary state at long times.

*Generalized Fokker-Planck equations in phase space.* As far as the central limit theorem or Pawula's theorem [34.35] is concerned, the statistical properties of the random force $L(t)$ in Eq. (12) turn out to be characterized only by its mean $\langle L(t_1) \rangle$ and its autocorrelation function $\langle L(t_1)L(t_2) \rangle$. Accordingly, in this Gaussian approximation the stochastic differential equations (11) are dubbed Langevin equations, whereas the Kolmogorov equation (12) reduces to the generalized Fokker-Planck equation, i.e.,

$$\frac{\partial \mathcal{F}(x,p,t)}{\partial t} = -\frac{p}{m}\frac{\partial \mathcal{F}(x,p,t)}{\partial x} + \frac{\partial}{\partial p}\left[\frac{\partial \mathcal{V}_{\text{eff}}(x,t)}{\partial x} + \beta p\right]\mathcal{F}(x,p,t) + \mathcal{D}_p(t)\frac{\partial^2 \mathcal{F}(x,p,t)}{\partial p^2}, \tag{21}$$

where $\mathcal{V}_{\text{eff}}(x,t)$ is an effective potential given by

$$\mathcal{V}_{\text{eff}}(x,t) = V(x) - xb\langle\Psi(t)\rangle, \tag{22}$$

$b\langle\Psi(t)\rangle$ being the average of the Langevin random force $L(t) = b\Psi(t)$, i.e.,

$$b\langle\Psi(t)\rangle = b\int_{-\infty}^{\infty} \psi \mathcal{F}_\Psi(\psi,t)d\psi, \tag{23}$$

where we have used the fact that $\langle\Psi(t)\rangle = \lim_{\epsilon\to 0}(1/\epsilon)\int_t^{t+\epsilon}\langle\Psi(t')\rangle dt'$. The novelty in Eq. (21) is the presence of the time-dependent diffusion coefficient $\mathcal{D}_p(t)$ given by

$$\mathcal{D}_p(t) = \beta m \mathcal{E}(t), \tag{24}$$

which is deemed to be defined in terms of the time-dependent diffusion energy

$$\mathcal{E}(t) = \mathcal{E}I(t), \tag{25}$$

where $I(t)$ is the dimensionless function

$$I(t) = \lim_{\epsilon\to 0}\frac{1}{\epsilon}\int_t^{t+\epsilon}\int_t^{t+\epsilon}\langle\Psi(t')\Psi(t'')\rangle dt'dt'', \tag{26}$$

and $\mathcal{E}$ the time-independent diffusion energy

$$\mathcal{E} = \frac{b^2}{2\beta m}, \tag{27}$$

yielding in turn the fluctuation-dissipation relation in the form

$$b = \sqrt{2\beta m \mathcal{E}}. \tag{28}$$

So the parameter $b$ measuring the fluctuation strength in the Langevin equation (11a) is determined by the friction constant $\beta$, the particle's mass $m$, as well as the diffusion energy $\mathcal{E}$.

The characteristic feature underlying the concept of time-dependent diffusion energy (25), with Eq. (27), i.e., $\mathcal{E}(t) = b^2 I(t)/2\beta m$, is that it sets up a general relationship between both fluctuation and dissipation phenomena: While the term $b^2 I(t)$ comes from the autocorrelation of the Langevin fluctuating force $L(t) = b\Psi(t)$, the frictional constant $\beta$ gives rise to dissipation processes.

For $I(t) = 0$, the generalized Fokker-Planck equation (21) reduces to the dissipative Liouville equation

$$\frac{\partial \mathcal{F}(x,p,t)}{\partial t} = -\frac{p}{m}\frac{\partial \mathcal{F}(x,p,t)}{\partial x} + \frac{\partial}{\partial p}\left[\frac{dV(x)}{dx} + \beta p\right]\mathcal{F}(x,p,t). \qquad (29)$$

*The generalized Klein-Kramers equation.* Let us consider an environment in thermodynamic equilibrium at temperature $T$ and characterized by the thermal energy $k_B T$, where $k_B$ denotes the Boltzmann constant. Upon identifying the Brownian particle's diffusion energy $\mathcal{E}$ with the reservoir's thermal energy $k_B T$, i.e.,

$$\mathcal{E} \equiv k_B T, \qquad (30)$$

the Fokker-Planck equation (21), with Eq. (24) and (25), reads

$$\frac{\partial \mathcal{F}(x,p,t)}{\partial t} = -\frac{p}{m}\frac{\partial \mathcal{F}(x,p,t)}{\partial x} + \frac{\partial}{\partial p}\left[\frac{\partial \mathcal{V}_{\text{eff}}(x,t)}{\partial x} + \beta p\right]\mathcal{F}(x,p,t) + \beta m k_B T I(t) \frac{\partial^2 \mathcal{F}(x,p,t)}{\partial p^2}. \qquad (31)$$

Under both conditions $\langle \Psi(t) \rangle = 0$ and $I(t) = 1$, and using $\beta \equiv 2\gamma$, the usual Klein-Kramers equation (1) is obtained as a special case. For this reason, we dub Eq. (31), for $I(t) \neq 1$, the generalized Klein-Kramers equation.

By integrating Eq. (31) for a free particle, $V(x) = 0$, over $x$ the generalized Rayleigh equation is obtained

$$\frac{\partial \mathcal{F}(p,t)}{\partial t} = -\frac{p}{m}\frac{\partial \mathcal{F}(p,t)}{\partial x} + \frac{\partial}{\partial p}[-b\langle \Psi(t) \rangle + \beta p]\mathcal{F}(p,t) + \beta m k_B T I(t) \frac{\partial^2 \mathcal{F}(p,t)}{\partial p^2}, (32)$$

Both conditions $\langle \Psi(t) \rangle = 0$ and $I(t) = 1$ lead to the Rayleigh equation (4).

*The correlational function $I(t)$.* For long times $t \to \infty$, if the correlational function (26) displays the following steady behavior

$$\lim_{t \to \infty} I(t) = 1, \qquad (33)$$

then we say that the fluctuations render Markovian in the sense that the time-dependence for the statistical autocorrelations of the Langevin force is "forgotten" in the steady regime. Accordingly, the particle-environment interaction is said to be non-Markovian in the non-steady range $0 < t < \infty$. According to that definition of Markovian and non-Markovian processes, the Fokker-Planck equations (1), (4) and (6), in which $I(t) = 1$, are Markovian, as expected, while the generalized Fokker-Planck equations (21), (31) and (32) are non-Markovian for $I(t) \neq 1$.

The non-Markovian character is also manifest if $I(t)$ does feature the following asymptotic behavior

$$\lim_{t \to \infty} I(t) \approx I'(t). \qquad (34)$$

In that case the stochastic process remains ever non-Markovian.

One example of $I(t)$ has been built up in Appendix B as

$$I(t) = \lambda \frac{t^{\lambda-1}}{t_c^{\lambda-1}} + 1 - e^{\frac{-t}{t_c}}, \qquad (35)$$

where the correlation time $t_c$ explicitly represents a non-Markovianity parameter between Brownian particle and environment. For $\lambda = 0$ or $\lambda < 1$ Expression (35) exhibits the stationary behavior (33) when $t \gg t_c$, whereas for $\lambda > 1$, as it will be shown in Sect. 4, the correlational function (35) leads to the anomalous diffusion phenomenon at large times.

## 3. Generalized Smoluchowski equation

As the inertial force is too small in comparison with the friction force, i.e., $|dP(t)/dt| \ll |\beta P(t)|$, the set of stochastic differential equations (11) satisfying the fluctuation-dissipation relation (27) reduces to

$$\frac{dX(t)}{dt} = -\frac{1}{\beta m}\frac{dV(X)}{dX} + \sqrt{\frac{2\mathcal{E}}{\beta m}}\Psi(t) \tag{36}$$

which in turn gives rise to the Kolmogorov stochastic equation in configuration space (see Appendix A)

$$\frac{\partial \mathcal{F}(x,t)}{\partial t} = \sum_{k=1}^{\infty} \frac{(-1)^k}{k!} \frac{\partial^k}{\partial x^k}[A_k(x,t)\mathcal{F}(x,t)], \tag{37}$$

the coefficients $A_k(x,t)$ being given by

$$A_k(x,t) = \lim_{\epsilon \to 0}\left[\frac{\langle \Delta X^k \rangle}{\epsilon}\right], \tag{38}$$

where the average values $\langle \Delta X^k \rangle$ are to be calculated about the sharp values $x'$, i.e.,

$$\mathcal{F}_{X\Psi}(x,\psi,t) = \delta(x - x')\mathcal{F}_{\Psi}(\psi,t). \tag{39}$$

*The non-Markovian Fokker-Planck equation in configuration space.* The Kolmogorov equation (37) in the Gaussian approximation reduces to the so-called non-Markovian Fokker-Planck equation in configuration space

$$\frac{\partial \mathcal{F}(x,t)}{\partial t} = \frac{1}{\beta m}\frac{\partial}{\partial x}\left[\frac{dV(x)}{dx}\mathcal{F}(x,t)\right] + \mathcal{D}_x(t)\frac{\partial^2 \mathcal{F}(x,t)}{\partial x^2}, \tag{40}$$

where we used $\langle \Psi(t) \rangle = 0$. In Eq. (40) $\mathcal{D}_x(t)$ denotes the time-dependent diffusion coefficient

$$\mathcal{D}_x(t) = \frac{\mathcal{E}}{\beta m}I(t), \tag{41}$$

associated with the motion of $\mathcal{F}(x,t)$. The diffusion energy $\mathcal{E}$ and the correlational function $I(t)$ in Eq. (41) are given by Eqs. (25) and (26), respectively.

*The non-Markovian Smoluchowski equation.* As far as $\mathcal{E} \equiv k_B T$ and $\beta \equiv \gamma$ are concerned, the Fokker-Planck equation (40) is called the non-Markovian Smoluchowski equation [31-33]

$$\frac{\partial \mathcal{F}(x,t)}{\partial t} = \frac{1}{\gamma m}\frac{\partial}{\partial x}\left[\frac{dV(x)}{dx}\mathcal{F}(x,t)\right] + \frac{k_B T}{\gamma m}I(t)\frac{\partial^2 \mathcal{F}(x,t)}{\partial x^2} \tag{42}$$

since for $I(t) = 1$ it is identical to the Markovian Smoluchowski equation (6).

# 4. Anomalous diffusion

In this section our aim is to show how our correlational function $I(t)$, Eq. (26), can be viewed as a physical mechanism giving rise to the law of anomalous diffusion (10). Initially, the anomalous diffusion phenomenon is studied for a free particle and for a harmonic oscillator in absence of inertial force, next it is looked at for a free particle in presence of such a force.

## 4.1. Anomalous diffusion in the absence of inertial force

*Free particle.* We begin with the Brownian motion of a free particle in absence of inertial force described by the Langevin equation (36), with $V(X) = 0$, i. e.,

$$\frac{dX(t)}{dt} = \sqrt{\frac{2\mathcal{E}}{\gamma m}}\,\Psi(t). \tag{43}$$

The corresponding diffusion equation (40), with Eq. (35), reads

$$\frac{\partial f(x,t)}{\partial t} = D_x^{(\lambda)}(t)\frac{\partial^2 f(x,t)}{\partial x^2}, \tag{44}$$

where $D_x^{(\lambda)}(t)$ is the time-dependent diffusion coefficient

$$D_x^{(\lambda)}(t) = \frac{\mathcal{E}}{\gamma m}\left(\lambda\frac{t^{\lambda-1}}{t_c^{\lambda-1}} + 1 - e^{\frac{-t}{t_c}}\right), \tag{45}$$

indexed by the parameter $\lambda > 1$ (see Appendix B). Starting from the initial condition $f(x, t = 0) = \delta(x)$, solution to Eq. (44) is given by the Gaussian function

$$f(x,t) = \frac{1}{\sqrt{2\pi\langle X^2(t)\rangle}}e^{\frac{-x^2}{2\langle X^2(t)\rangle}}, \tag{46}$$

expressed in terms of the mean square displacement

$$\langle X^2(t)\rangle = \frac{2\mathcal{E}}{\gamma m}\left[\frac{t^\lambda}{t_c^{\lambda-1}} + t + t_c\left(e^{\frac{-t}{t_c}} - 1\right)\right], \tag{47}$$

which displays the following anomalous behavior at long-time $t \to \infty$

$$\langle X^2(t)\rangle \sim \frac{2\mathcal{E}}{\gamma m t_c^{\lambda-1}}t^\lambda, \qquad \lambda > 1. \tag{48}$$

So it is predicted that the anomalous Brownian motion of a free particle immersed in a generic environment characterized by the diffusion energy $\mathcal{E}$ is superdiffusive and non-Markovian. In the case of a thermal bath, we have $\mathcal{E} \equiv k_B T$.

*Harmonic oscillator.* The non-Markovian Smoluchowski equation (40), with Eq. (35), reads

$$\frac{\partial f(x,t)}{\partial t} = \frac{\omega^2}{\gamma}\frac{\partial}{\partial x}[xf(x,t)] + \frac{\mathcal{E}}{\gamma m}\left(\lambda \frac{t^{\lambda-1}}{t_c^{\lambda-1}} + 1 - e^{\frac{-t}{t_c}}\right)\frac{\partial^2 f(x,t)}{\partial x^2}, \qquad (49)$$

(i) For $\lambda = 2$, we find

$$\langle X^2(t)\rangle = \frac{\mathcal{E}}{m\omega^2}\left\{\frac{2}{t_c}\left[t + \tau\left(e^{\frac{-t}{\tau}} - 1\right)\right] + 1 - e^{\frac{-t}{\tau}} - \frac{t_c}{(t_c - \tau)}\left(e^{\frac{-t}{t_c}} - e^{\frac{-t}{\tau}}\right)\right\} \qquad (50)$$

and the Einsteinian diffusion law

$$\langle X^2(t)\rangle \sim \frac{2\mathcal{E}}{m\omega^2 t_c} t, \quad t \to \infty; \qquad (51)$$

(ii) for $\lambda = 3$, we have

$$\langle X^2(t)\rangle = \frac{\mathcal{E}}{m\omega^2}\left\{\frac{3}{t_c^2}\left[t^2 - 2t\tau + 2\tau^2\left(1 - e^{\frac{-t}{\tau}}\right)\right] + 1 - e^{\frac{-t}{\tau}} - \frac{t_c}{(t_c - \tau)}\left(e^{\frac{-t}{t_c}} - e^{\frac{-t}{\tau}}\right)\right\} \qquad (52)$$

and the ballistic diffusion

$$\langle X^2(t)\rangle \sim \frac{3\mathcal{E}}{m\omega^2 t_c^2} t^2, \quad t \to \infty, \qquad (53)$$

(iii) for $\lambda = 4$, Eq. (49) leads to

$$\langle X^2(t)\rangle = \frac{\mathcal{E}}{m\omega^2}\left\{\frac{4}{t_c^3}\left[t^3 - 3\tau t^2 + 6\tau^2 t - 6\tau^3\left(1 - e^{\frac{-t}{\tau}}\right)\right] + 1 - e^{\frac{-t}{\tau}}\right.$$
$$\left. - \frac{t_c}{(t_c - \tau)}\left(e^{\frac{-t}{t_c}} - e^{\frac{-t}{\tau}}\right)\right\} \qquad (54)$$

and the turbulent diffusion

$$\langle X^2(t)\rangle \sim \frac{4\mathcal{E}}{m\omega^2 t_c^3} t^3, \quad t \to \infty, \qquad (55)$$

In Eqs. (50,52,54), $\tau$ denotes the relaxation time $\tau = \gamma/2\omega^2$. By generalizing results (51), (53) and (55) the anomalous behavior reads

$$\langle X^2(t)\rangle \sim \lambda \frac{\mathcal{E}}{m\omega^2} \frac{t^{\lambda-1}}{t_c^{\lambda-1}}, \quad \text{for } \lambda \geq 3. \tag{56}$$

The anomalous Brownian motion of a harmonic oscillator immersed in a generic environment $\mathcal{E}$ is superdiffusive and non-Markovian. If the environment is a thermal bath, then $\mathcal{E} \equiv k_B T$.

### 4.2 Anomalous diffusion in the presence of inertial force

*Anomalous Brownian motion in momentum space.* In order to examine anomalous Brownian motion of a free particle in momentum space we integrate the Fokker-Planck equation (21) over $x$ and obtain the generalized Rayleigh equation

$$\frac{\partial \mathcal{F}(p,t)}{\partial t} = 2\gamma \frac{\partial}{\partial p}[p\mathcal{F}(p,t)] + 2\gamma m \mathcal{E}\left(\lambda \frac{t^{\lambda-1}}{t_c^{\lambda-1}} + 1 - e^{\frac{-t}{t_c}}\right)\frac{\partial^2 \mathcal{F}(p,t)}{\partial p^2}, \tag{57}$$

where we used the correlational function (35).

We take into account the following cases:

(i) for $\lambda = 2$, we find the mean square momentum

$$\langle P^2(t)\rangle = m\mathcal{E}\left\{\frac{2}{t_c}\left[t + \frac{1}{4\gamma}(e^{-4\gamma t} - 1)\right] + 1 - e^{-4\gamma t} + \frac{4\gamma t_c}{4\gamma t_c - 1}\left(e^{-4\gamma t} - e^{\frac{-t}{t_c}}\right)\right\}, \tag{58}$$

exhibiting the Einsteinian diffusive regime

$$\langle P^2(t)\rangle \sim \frac{2m\mathcal{E}}{t_c} t, \quad t \to \infty, \tag{59}$$

(ii) for $\lambda = 3$, we have

$$\langle P^2(t)\rangle = m\mathcal{E}\left\{\frac{3}{t_c^2}\left[t^2 - \frac{t}{2\gamma} + \frac{2}{(4\gamma)^2}(1 - e^{-4\gamma t})\right] + 1 - e^{-4\gamma t}\right.$$
$$\left. + \frac{4\gamma t_c}{4\gamma t_c - 1}\left(e^{-4\gamma t} - e^{\frac{-t}{t_c}}\right)\right\} \tag{60}$$

displaying the following ballistic behavior

$$\langle P^2(t)\rangle \sim \frac{3m\mathcal{E}}{t_c^2} t^2, \quad t \to \infty; \tag{61}$$

(iii) for $\lambda = 4$, Eq. (55) yields

$$\langle P^2(t) \rangle = m\mathcal{E}\left\{\frac{16\gamma}{t_c^3}\left[\frac{t^3}{4\gamma} - \frac{3t^2}{(4\gamma)^2} + \frac{6t}{(4\gamma)^3} - \frac{6(1-e^{-4\gamma t})}{(4\gamma)^4}\right] + 1 - e^{-4\gamma t} \right. \\ \left. + \frac{4\gamma t_c}{4\gamma t_c - 1}\left(e^{-4\gamma t} - e^{\frac{-t}{t_c}}\right)\right\} \tag{62}$$

leading to the turbulent regime

$$\langle P^2(t) \rangle \sim \frac{4m\mathcal{E}}{t_c^3}t^3, \quad t \to \infty, \tag{63}$$

Generalizing the long-time results (59), (61) and (63) leads to the anomalous diffusion in momentum space

$$\langle P^2(t) \rangle \sim \lambda m\mathcal{E}\frac{t^{\lambda-1}}{t_c^{\lambda-1}}, \quad \text{for } \lambda \geq 3. \tag{64}$$

In brief, the non-Markovian diffusion of the mean square momentum of a Brownian free particle in presence of inertial force is superdiffusive and anomalous for $\lambda \geq 3$.

## 5. Summary and concluding remarks

In this paper we have studied the anomalous (or non-Einsteinian) diffusion of a Brownian particle immersed in a generic environment within the framework of generalized linear Fokker-Planck equations (21) and (40). Our main findings are the non-Markovian quantities (48), (56), and (64) on the basis of which some theoretical predictions were the following:

i) the anomalous diffusion of a free particle and a harmonic oscillator, described by the mean square displacement (48) and (56), respectively, is superdiffusive in absence of inertial force;

ii) the anomalous diffusion of a free particle, given by the mean square momentum (64), is superdiffusive in presence of inertial force.

It is worth noticing that current approaches have been developed to tackle the anomalous diffusion phenomenon either via generalized entropies generating nonlinear Fokker-Planck equations or through fractional Fokker-Planck equations. In contrast to such approaches, the present study showed that anomalous diffusion could be investigated within the framework of linear Fokker-Planck equations.

From a perspective of generalized statistical mechanics the following question can be raised on basis of our non-Markovian approach to anomalous diffusion: What are the non-Boltzmannian entropies to be maximized by the time-dependent solutions to our non-Markovian linear Fokker-Planck equations (44), (49), and (57)? A possible answer to that question could point out the relevance of stochastic dynamics for deriving entropic forms.

# Appendix A. Kolmogorov equations

We suppose the dynamics of the stochastic process $\Phi = \Phi(t)$ to be governed by the ordinary differential equation

$$\frac{d\Phi(t)}{dt} = \mathcal{K}(\Phi, t). \qquad (A.1)$$

To find out the time evolution of the probability distribution function $\mathcal{F}(\varphi, t)$ expressed in terms of the realizations $\varphi$ of the random variable $\Phi$, we closely follow Stratonovich's procedure [31-33,36]. We resort to the definition of conditional probability density given by

$$W(\varphi', t'|\varphi, t) = \frac{f(\varphi', t'; \varphi, t)}{\mathcal{F}(\varphi, t)}, \qquad (A.2)$$

where $f(\varphi, t; \varphi', t')$ is the joint probability density function of $\Phi$ at different times $t$ and $t'$ with $\varphi' \equiv \varphi(t')$ and $\varphi \equiv \varphi(t)$.

From Eq. (A.2), we arrive at the Bachelier-Einstein integral equation [6, 37]

$$\mathcal{F}(\varphi', t') = \int_{-\infty}^{\infty} W(\varphi', t'|\varphi, t)\mathcal{F}(\varphi, t)d\varphi, \qquad (A.3)$$

after using the Kolmogorov compatibility condition [38]

$$\int_{-\infty}^{\infty} f(\varphi', t'; \varphi, t)d\varphi = \mathcal{F}(\varphi', t'). \qquad (A.4)$$

The characteristic function for the increment $\Delta\Phi \equiv \Phi(t') - \Phi(t)$ is expressed in terms of the conditional probability density (A.2) as

$$\langle e^{iu\Delta\Phi} \rangle = \int_{-\infty}^{\infty} e^{iu\Delta\varphi} W(\varphi', t'|\varphi, t)d\varphi, \qquad (A.5)$$

the inverse of which is

$$W(\varphi', t'|\varphi, t) = \frac{1}{2\pi} \int_{-\infty}^{\infty} e^{-iu\Delta\varphi} \langle e^{iu\Delta\Phi} \rangle du. \qquad (A.6)$$

Now, using the expansion

$$\langle e^{iu\Delta\Phi}\rangle = \sum_{s=0}^{\infty} \frac{(iu)^s}{s!} \langle(\Delta\Phi)^s\rangle, \quad (A.7)$$

Eq. (A.6) turns out to be given by

$$W(\varphi',t'|\varphi,t) = \sum_{s=0}^{\infty} \frac{(-1)^s}{s!} \langle(\Delta\Phi)^s\rangle \frac{\partial^s}{\partial\varphi^s} \delta(\varphi-\varphi'). \quad (A.8)$$

Inserting (A.8) into (A.3) and dividing the resulting equation by $\epsilon$, we obtain

$$\frac{\mathcal{F}(\varphi',t') - \mathcal{F}(\varphi',t)}{\epsilon} = \sum_{s=1}^{\infty} \frac{(-1)^s}{s!} \frac{\partial^s}{\partial\varphi^s} \left\{ \frac{\langle(\Delta\Phi)^s\rangle}{\epsilon} \mathcal{F}(\varphi,t) \right\}, \quad (A.9)$$

with $t' = t + \epsilon$ and $\varphi' \equiv \varphi(t+\epsilon)$. The procedure of taking the limit $\epsilon \to 0$ in both sides of (A.9) leads to the Kolmogorov equation

$$\frac{\partial \mathcal{F}(\varphi,t)}{\partial t} = \mathbb{K}\mathcal{F}(\varphi,t), \quad (A.10)$$

where the Kolmogorovian operator $\mathbb{K}$ acts upon the probability distribution function $\mathcal{F}(\varphi,t)$ according to

$$\mathbb{K}\mathcal{F}(\varphi,t) = \sum_{s=1}^{\infty} \frac{(-1)^s}{s!} \frac{\partial^s}{\partial\varphi^s} [B^{(s)}(\varphi,t)\mathcal{F}(\varphi,t)], \quad (A.11)$$

with the coefficients $B^{(s)}(\varphi,t)$, given by

$$B^{(s)}(\varphi,t) = \lim_{\epsilon \to 0} \frac{\langle(\Delta\Phi)^s\rangle}{\epsilon}, \quad (A.12)$$

calculated from the stochastic differential equation (A.1) in the following integral form

$$\Delta\Phi \equiv \Phi(t+\epsilon) - \Phi(t) = \int_{t}^{t+\epsilon} \mathcal{K}(\Phi,t')dt', \quad (A.13)$$

after averaging $(\Delta\varPhi)^s$ over a given conditional probability density $W(\varphi',t'|\varphi,t)$ according to Eq. (A.5).

Summing up, we have set out a general scheme for deriving from the stochastic differential equation (A.1) the Kolmogorov equation (A.10) which reckons with non-Gaussian features on account of the presence of the $s$th moment of the increment $\Delta\varPhi$, i.e., $\langle(\Delta\varPhi)^s\rangle$.

According to Pawula's theorem [34,35], there exists no non-Gaussian approximation to the non-Gaussian Kolmogorov equation (A.10) in compliance with the positivity of $\mathcal{F}(\varphi,t)$. Hence, in the Gaussian approximation Eq. (A.10) reads

$$\frac{\partial \mathcal{F}_\varPhi(\varphi,t)}{\partial t} = -\frac{\partial}{\partial\varphi}\big[B^{(1)}(\varphi,t)\mathcal{F}_\varPhi(\varphi,t)\big] + \frac{1}{2}\frac{\partial^2}{\partial\varphi^2}\big[B^{(2)}(\varphi,t)\mathcal{F}_\varPhi(\varphi,t)\big]. \quad (A.14)$$

In the physics literature, the class of stochastic differential equation (A.1) in the Gaussian approximation is known as Langevin equation (e. g., Eqs. (11) and (36)), whereas the corresponding Gaussian Kolmogorov equation (A.14) is dubbed Fokker-Planck equation in configuration space or in momentum space, for example, the non-Markovian Smoluchowski equation (32) and the non-Markovian Rayleigh equation (42), respectively.

For the case of two random variables $\varPi$ and $\varPhi$, the set of stochastic differential equations, given by

$$\frac{d\varPi(t)}{dt} = \mathcal{K}_1(\varPi,\varPhi,t), \quad (A.15a)$$

$$\frac{d\varPhi(t)}{dt} = \mathcal{K}_2(\varPi,\varPhi,t), \quad (A.15b)$$

generates the following phase space Kolmogorov equation for the joint probability distribution $\mathcal{F}_{\varPi\varPhi}(\pi,\varphi,t)$

$$\frac{\partial \mathcal{F}_{\varPi\varPhi}(\pi,\varphi,t)}{\partial t} = \sum_{s=1}^{\infty}\sum_{r=0}^{s}\frac{(-1)^s}{r!(s-r)!}\frac{\partial^s}{\partial\pi^{s-r}\partial\varphi^r}\big[B^{(s-r,r)}(\pi,\varphi,t)\mathcal{F}_{\varPi\varPhi}(\pi,\varphi,t)\big], \quad (A.16)$$

with

$$B^{(s-r,r)}(\pi,\varphi,t) = \lim_{\epsilon\to 0}\left[\frac{\langle(\Delta\Pi)^{s-r}(\Delta\Phi)^r\rangle}{\epsilon}\right]. \qquad (A.17)$$

The increments $\Delta\Pi$ and $\Delta\Phi$ are evaluated from both Eqs. (A.15) in the integral form

$$\Delta\Phi \equiv \Phi(t+\epsilon) - \Phi(t) = \int_{t}^{t+\epsilon} \mathcal{K}_1(\Pi,\Phi,t)dt \qquad (A.18)$$

and

$$\Delta\Pi \equiv \Pi(t+\epsilon) - \Pi(t) = \int_{t}^{t+\epsilon} \mathcal{K}_2(\Pi,\Phi,t)dt. \qquad (A.19)$$

In the Gaussian approximation the phase space Kolmogorov equation (A.16) reduces to the following Fokker-Planck equation in phase space

$$\frac{\partial \mathcal{F}_{\Pi\Phi}(\pi,\varphi,t)}{\partial t} = \sum_{s=1}^{2}\sum_{r=0}^{s}\frac{(-1)^s}{r!(s-r)!}\frac{\partial^s}{\partial \pi^{s-r}\partial \varphi^r}\left[B^{(s-r,r)}(\pi,\varphi,t)\mathcal{F}_{\Pi\Phi}(\pi,\varphi,t)\right]. \quad (A.20)$$

Both Langevin equations (11) in phase space are physical examples of equations of motion (A.15), while the non-Markovian Klein-Kramers equation (31) is a special case of phase space Fokker-Planck equation (A.20).

In Ref. [31,33] we have found out explicitly the coefficients $B^{(s)}(\varphi,t)$ in Eq. (A.14) for the cases of the non-Markovian Rayleigh equation (32) and the non-Markovian Smoluchowski equation (42), as well as the coefficients $B^{(s-r,r)}(\pi,\varphi,t)$ concerning the non-Markovian Klein-Kramers equation (31).

# Appendix B. The correlational function $I(t)$

The correlational function $I(t)$ is defined in Eq. (26) as

$$I(t) = \lim_{\epsilon \to 0} \frac{1}{\epsilon} \int_t^{t+\epsilon} \int_t^{t+\epsilon} \langle \Psi(t')\Psi(t'')\rangle dt'dt'' \geq 0. \qquad (B.1)$$

If the autocorrelation function $\langle \Psi(t')\Psi(t'')\rangle$ is given by

$$\langle \Psi(t')\Psi(t'')\rangle = \left(\lambda \frac{t'^{\lambda-1}}{t_c^{\lambda-1}} + 1 - e^{\frac{-(t'+t'')}{2t_c}}\right)\delta(t'-t''), \qquad (B.2)$$

where $t_c$ is deemed to be the correlation time of $\Psi(t)$ at times $t'$ and $t''$, then it follows that

$$I(t) = \lambda \frac{t^{\lambda-1}}{t_c^{\lambda-1}} + 1 - e^{\frac{-t}{t_c}}. \qquad (B.3)$$

Function (B.3) displays the following asymptotic behavior at long times $t \to \infty$:

$$I(t) \sim \lambda \frac{t^{\lambda-1}}{t_c^{\lambda-1}}, \qquad (B.4)$$

as long as $\lambda > 1$.